# The Factors of Code Reviewing Process to Ensure Software Quality

Shaykh Siddique
Department of Computer Science and Engineering,
East West University
Dhaka, Bangladesh
shaykhsiddiqee@gmail.com

*Abstract*— In the era of revolution, the development of softwares are increasing daily. The quality of software impacts the most in software development. To ensure the quality of the software it needs to be reviewed and updated. The effectiveness of the code review is that it ensures the quality of software and makes it updated. Code review is the best process that helps the developers to develop a system errorless. This report contains two different code review papers to be evaluated and find the influences that can affect the code reviewing process. The reader can easily understand the factor of the code review process which is directly associated with software quality assurance.

## I. INTRODUCTION

### A. Background

To ensure the proper quality of software, many monitoring methods are necessary. Different kinds of software (e.g. web application, computer application, Mobile application, games) need various kinds of testing and judging. Generally, factors and model analysis of software [1], measuring metrics [2], inspection [3], reachability analysis [4], cyclomatic complexity analysis [5] all are different steps of assuring the quality of software. The code reviewing process (CRP) [6] can help to improve software quality while developing a software module as well as that is already developed. This technique is a white box testing technique because the testing engineer has to go through the source code.

Code Review Process is very popular in the sector of code investigation. Software Quality Assurance has so many fields, where CRP is a very principal. Software industries mostly dependent on code reviews because it meets both ends of customers requirements. Code review can be individual peer work or it can be a group or pair work. As industry wants to develop software based on users' requirements, developers need to rework and update their codes regularly. Code review is very essential to resolve this issue.

The aim of the study is to investigate the factors and the influences of the code review process to ensure software quality. The research objectives are:
1. To conduct of effective code review process.
2. To analyze the influencing factors of the code review process for software quality assurance.

This paper has been organized into six sections. Section 1 defines the introduction and problem description. The literature review of two previous studies are reported in section 2. In the third part, the methodology is defined. Section 4 contains the result part of our objectives. The discussion and conclusion are added in section 5 and section 6.

The main research study question is, what are the factors influences the code reviewing techniques to assure software quality. Also, have some sub-questions:
1. How a new reviewer can gain knowledge about the code review process?
2. What are the factors of the code review process at the industry level?

## II. LITERATURE REVIEW

The large number of work are done which are related to modern code review systems, technique, and tools. German researcher Tobias Baum found the major factors influencing the code review process for the industry level [7]. They interviewed software development professionals from 19 different companies. Those provided the influences and the reasons for adopting a different code review process to improve Software quality. The desired effects they found for code reviews were- better code quality, finding defects, learning (reviewers and authors), sense of mutual responsibility, finding better solutions, and complying with QA guidelines. The undesired effects were- stuff efforts, increased life cycle time, and Offending the author. They also generated some hypothesis as, "Code review processes are mainly introduced when a gap between some goal and reality, is perceived", "Code review is most likely to remain in use if it is embedded into the process so that it does not require a conscious decision to do a review". The main factors shaping the review process came from their research. The code review process is influenced by the process used by role models, used review tools, culture, development team, product, and development process. The sources of learning code review process ware from colleagues or other teams in the same company, open-source projects, blogs, and web pages, university education, practitioners journals and conferences, and books on software engineering. The limitations of their study were, as they used the interview, the data could be biased.

Another researcher Oleksii Kononenko [8] surveyed the code review process and how developers see it. They summarized the survey results of 88 Mozilla core developers. The results defined the review quality, how they evaluate the submitted code, and what challenges they face when performing the code review process. The qualitative analysis of their research aims about- how do Mozilla developers conduct code review and also what factors developers consider for code review time and decision. Their works also imply the factors that developers use to assess code review quality. According to their study, the reviewer often finds difficulties to keep their technical skills up-to-date, manage personal priorities, and mitigate context switching and the factors that included technical, personal, and social signals to contribute to the review quality. Their survey consisted of three major parts: nine questions about the developer's background and real working, three questions related to a different feature of code review, and seven open-ended questions on issues raised by the multiple-choice questions on code review system which takes 5-10 minutes to participants. All of their potential participants were at least 3 years industrial experienced as the code reviewer. The grounded theory methodology to analyze their dataset. Patch size (LOC) was the factor influencing code review time for more than 90% of participants, reviewer experience, and number of modified file effects for 70% participants, code chunks, patch writer experience, number of resubmits, review queue were also positive effects on their results. During open-end coding code quality, testing, time constraints, change scope/rationale, nature of the bug, understanding code change/base all are responded as different kinds of parameters. Perception of patch quality, characteristics of a well-done code review assessed the code review quality. About 95% of the participants were agreed that the reviewer experience can produce a better code review process quality. 76% to 85% of participants also defines the personal factors such as patch writer experience, reviewer workloads, developer participation in the discussion of code changes, number of resubmitted are more likely to affect the quality of reviews. When the participants were asked about the challenges they faced for code reviewing process, most of then said about the technical challenges, gaining familiarity with the code, coping with the code complexity, and having suitable tool support. The tool they mostly used was Bugzila [9], Mozilla bug tracking, and management tool. As they surveyed on participants, so the dataset could be biased which they included as their limitation. Also, the analysis of data, data cleaning, and splitting could be biased also.

### III. METHODOLOGY

As the study is focused on the influences and the affects of the code review process (CRP), observation of theoretical methodology is used to find the results. Two research papers were reviewed and analyzed to find the possible outcome [7][8]. The correlations between the two studies' results were compared. For representing the diagrams and graphs we used MS Word drawing tools.

### IV. RESULTS

#### A. Conducting The Effective Code Reviews Process

Generally, reviewers are introduced about the code review process (CRP) from the following sources according to the analysis.
- Colleagues in the same company
- Blogs and web pages
- journals and conferences
- University Education
- Books related to SQA and CRP

The code review process also includes the following terms:

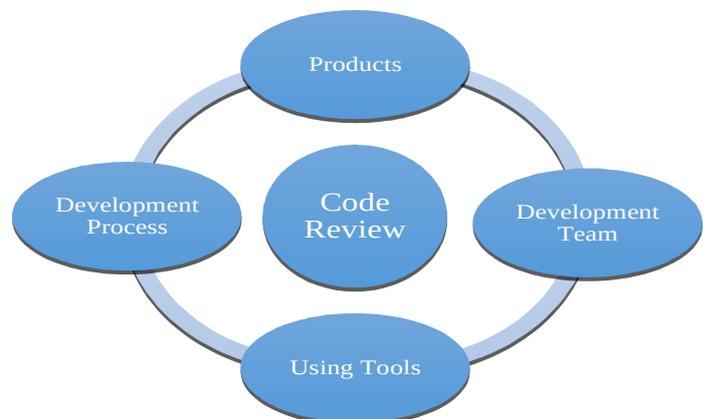

Figure 1: Planning for the code review process

Depend on the product, the author may choose the code review process. Also, the size of the development team can vary in the code review process. Chosen development process (e.g. agile, Waterfall, DevOps) the code review process can be adopted or not. The tools developers used to develop the software can be also included.

### B. Factors that influence the Code Review Process

There are many factors which are influence the code review process. The code review process extremely depends on the reviewers' experience. For the much-experienced reviewer, the quality improvement is obvious. As well as the patch size and the code chunks hold the second and third positions as influencer.

TABLE 1: FACTORS THAT INFLUENCE THE CODE REVIEW PROCESS

| Reviewer experience |
| --- |
| Patch size (LOC) |
| Code chunks |
| Number of modified files |
| Patch writer experience |
| Review queue |
| Module |
| Number of people in the discussion |
| Number of resubmits |
| Review response time |
| The length of the discussion |
| The severity of a bug |
| The priority of a bug |

The other factors are the number of modified files, review queue, the number of people in the discussion, resubmits, etc. which are included in Table. 1.

## V. DISCUSSION

To ensure the proper quality of software, the code review process plays a very important role. Our study defines some factors of the code review process. The result of conducting efficient code review which we found implies with the article on a website [10]. As well as the second result of our study about the factors that influence the code review process also match with another research study of Microsoft [11].

The code review process is a kind of white-box testing, which can be useful at any stage of development and testing. CRP can be used during unit testing, in integration testing, and also in the inspection.

## VI. CONCLUSION

Concluding a report indicates that the report contained all information successfully. This research report emphasized two previous papers and extract new information from them. The objectives to find influencing factors of the Code Review Process and conducting them in the effective Code Reviews have been successfully demonstrated in this report. Code review itself effectively influences the software industries. Finding the influence factors were very challenging also seems interesting. Effective Code Review can easily upgrade any system, for this Code Review needs proper inspection and documentation. This research report not only reviewed two papers but also extract a piece of information.